\documentclass{article}

\usepackage{microtype}
\usepackage{graphicx}
\usepackage{subfigure}
\usepackage{booktabs} 
\usepackage[T1]{fontenc}
\usepackage{mathtools}
\usepackage{amssymb}

\usepackage{hyperref}

\usepackage[accepted]{icml2024}

\usepackage{amsmath}
\usepackage{amssymb}
\usepackage{mathtools}
\usepackage{amsthm}

\usepackage[capitalize,noabbrev]{cleveref}

\theoremstyle{plain}

\theoremstyle{definition}

\theoremstyle{remark}

\usepackage[textsize=tiny]{todonotes}

\icmltitlerunning{Changepoint Detection in Highly-Attributed Dynamic Graphs}

\begin{document}

\twocolumn[
\icmltitle{Changepoint Detection in Highly-Attributed Dynamic Graphs}



\icmlsetsymbol{equal}{*}

\begin{icmlauthorlist}
\icmlauthor{Emiliano Penaloza}{equal,mila,montreal}\icmlauthor{ Nathaniel Stevens}{waterloo}
\end{icmlauthorlist}

\icmlaffiliation{mila}{Mila - Quebéc AI Institute}
\icmlaffiliation{montreal}{DIRO, Université de Montréal}
\icmlaffiliation{waterloo}{Department of Statistics \& Actuarial Science, University of Waterloo}

\icmlcorrespondingauthor{Emiliano Penaloza}{emilianopp550@gmail.com}

\icmlkeywords{Machine Learning, ICML}

\vskip 0.3in
]



\printAffiliationsAndNotice{}  

\begin{abstract}

Detecting anomalous behavior in dynamic networks remains a constant challenge. This problem is further exacerbated when the underlying topology of these networks is affected by individual highly-dimensional node attributes. We address this issue by tracking a network's modularity as a proxy of its community structure. We leverage Graph Neural Networks (GNNs) to estimate each snapshot's modularity. GNNs can account for both network structure and high-dimensional node attributes, providing a comprehensive approach for estimating network statistics. Our method is validated through simulations that demonstrate its ability to detect changes in highly-attributed networks by analyzing shifts in modularity. Moreover, we find our method is able to detect a real-world event within the \#Iran Twitter reply network, where each node has high-dimensional textual attributes. We additionally make our code available on Github \footnote{\href{https://github.com/Emilianopp/ChangesHighlyAttributedGraphs}{https://github.com/Emilianopp/ChangesHighlyAttributedGraphs}}.
\end{abstract}

\section{Introduction}
A network or graph is a collection of interconnected items, called nodes or vertices, found in areas like brain networks \cite{NetworksBrain}, as well as in human-engineered contexts such as computer networks  \cite{NetworksComputer}. Network science studies these structures, revealing their benefits in applications including social network mining \cite{SocialNetworks} and combatting human trafficking \cite{HumanTrafficking}. As the field evolved, subfields like network monitoring emerged, focusing on changes in network behaviour over time. Methods for detecting changes include statistical process monitoring and anomaly detection in time series used in engineering, manufacturing, and fraud detection \cite{manuControl,controlTextbook,eng}. Network monitoring often combines these methods, such as using statistical control charts or deep learning to detect anomalies in dynamic networks \cite{introCite}. Another subfield studies attributed networks, where nodes hold important information. Statistical inference methods are proficient at estimating parameters of the graph generation process but are largely limited by restrictive and unrealistic assumptions \cite{NewmanAttributed}.  On the other hand, GNNs have achieved state-of-the-art performance in classification, regression, and generative tasks over attributed and unattributed networks without such assumptions \cite{GraphSurvey}. Most network monitoring techniques ignore network attributes, focusing only on the network structure itself. Some works consider attributes but use simpler generalized linear models, limiting their capacity \cite{LogisticMonitoring,LogisticMonitoring2}. This study presents a method for monitoring networks with many attributes by tracking the graph's community structure. Communities are groups of similar, frequently interacting individuals. We find these groups by maximizing network modularity using a GNN, leveraging structural properties and high-dimensional attributes. When monitoring a time series of graphs, we detect structural changes when modularity changes significantly. Through simulations, we demonstrate that our method can detect a variety of changes in highly attributed networks. Additionally, we showcase our methods' practicality on a real world Twitter network, where the node attributes are derived from user tweets. We relegate the related work section to App. \ref{related_work}, where we note to the best of our knowledge this work is the first to study anomaly detection on \textit{highly}- attributed \textit{dynamic} graphs, specifically those with textual attributes. 

\section{Background}

\subsection{Dynamic Networks}
A dynamic network can be described as an ordered sequence of time varying graphs $\mathcal{G}(\mathcal{T}) = \{G_1,...,G_T\}$ over some timeframe $\mathcal{T} =\{ 1,...,T\}$. Each of these graphs may be independent, but most often they exhibit temporal dependence. We refer to these instances  $G_1,G_2,...,G_T$ as network snapshots. Each snapshot contains a set of nodes, $\mathcal{V}(t) = \{1,2,...,n_t\}$, edges $\mathcal{E}(t) = \{ e_{u,v}(t):u,v \in \mathcal{V}(t) , t \in \mathcal{T} \}$, and a set of $s$-dimensional attributes  $\mathcal{X}(t) = \{ x_u(t): u \in \mathcal{V}(t) , t \in \mathcal{T} \}$. It is important to note that given our definition of a dynamic network, the set of edges and nodes are expected to change between snapshots.  
Additionally, we define the neighborhood of first-degree connections of a node $u$ at time $t$ as $\mathcal{N}_u(t) = \{v:e_{u,v}(t) \in \mathcal{E}(\mathcal{T})\}$ at timestamp $t$. In this work, we encode each network snapshot using an adjacency matrix, denoted $\mathbf{A}_t$. The matrix can either have binary entries $A_{uv,t} = e_{u,v}(t) \in {0,1}$ or real-valued edge weights $A_{uv,t} = e_{u,v}(t) \in \mathbb{R}$. 

\subsection{ Deep Modularity Networks}
Traditionally, graph communities are obtained using self-supervised learning combined with k-means clustering on the derived node representations. However, these methods can be unreliable depending on the graph's size and sparsity \cite{kmeansCitation}. To address this, it is effective to directly optimize for a statistic that measures community structure. For example, modularity which measures the difference between the observed number of intra-cluster edges and what would be expected in a randomly generated graph.
\begin{align}
    Q &= \frac{1}{2w} \sum_{uv} \bigg [A_{uv} - \frac{d_ud_v}{2w} \bigg] \delta_{c_u,c_v}
\end{align}
where $\delta_{c_u,c_v}$ is the Kronecker delta, $w$ indicates the total number of edges in the graph, $d_u = |\mathcal{N}_{u}|$ is the degree of node $u$, and $c_u$ indicates the community assignment for node $u$. Unfortunately, the computation of the modularity gradient is intractable, making spectral modularity, a convenient alternative \cite{spectralMod}: 
\begin{align}
    Q = \frac{1}{2w} \text{Tr}(C^\top B C)
\end{align}
where $C \in \{0,1\}^{n\times k}$ is a community assignment matrix, $k$ is the number of communities, and $B = A - \frac{dd^\top}{2w}$ with $d\in \mathbb{R}^n$ being the degree vector. Traditionally, one can maximize the optimal modularity by choosing the assignment matrix $C$ to equal the top $k$ eigenvectors of $B$, but such a derivation is agnostic to node attributes. Recently, \citet{dmon} showed that a GNN  can be used to directly optimize spectral modularity and as a byproduct generate cluster assignments which can account for anode ttributes. Deep Modularity Networks (DMoN), achieve this by using a GNN to generate, a modified version of the community affiliation matrix, $\mathcal{C}\in \mathbb{R}^{{(0,1)}^{n \times k}}$, in spectral modularity:
\begin{align}
    \mathcal{C} = \text{softmax}(\text{GNN}(\tilde A , \mathcal{X} ))
\end{align}
where the GNN, usually a graph convolutional network \cite{gcn}, takes in the degree normalized adjacency matrix, $\tilde A =  D^{\frac{-1}{2}} (A) D^{\frac{-1}{2}}$. Given this, DMoN is able to directly optimize spectral modularity, through traditional gradient optimization, yielding a solution that accounts for both graph structure and attributes.

\begin{table*}[h]
\centering
\resizebox{.75\textwidth}{!}{%
\begin{tabular}{|c|c|c|c|}
\hline
 Change Type & Change Detection Percentage  &Conditional Expected Delay & Average Percentage Over Threshold \\ \hline
Community Splitting       & 1.0          & 1.0  & 1.0                                                                          \\ 
Community Merging       & 1.0          & 1.0  & 1.0\\
New Community       & 1.0          & 1.0  & 1.0    \\
Attribute Change      & 1.0          & 23.38  & 0.44   \\
\hline
\end{tabular}}
\caption{Performance summaries across $N$= 100 replications of the community splitting change}
\label{CommunitySplittingTable}
\end{table*}

\subsection{Methodology}
Since our goal is to identify changes in highly attributed \textit{dynamic} graphs, we can optimize community assignments using DMoN and directly track the associated modularity metric. Other than introducing DMoN, \citet{dmon} also introduced the collapse regularizer (CR) that penalizes trivial solutions that arise when optimizing spectral modularity:
\small
\begin{align}
    \text{CR} = \frac{\sqrt{k}}{n} \Bigg|\Bigg| \text{ColSum}(\mathcal{C})\Bigg|\Bigg|_2 -1 .
\end{align}
\normalsize

For example, this regularization penalizes the case for which all $k$ communities collapse into a single one by taking its maximum value at such a solution. While this regularizer provides stable solutions for cases when the model tries to collapse the clusters onto as single community, it fails to account for the trivial solution of assigning equal weight to all clusters for each node. We instead propose the square root collapse optimizer (SRCO) to address these limitations:
\small
\begin{align}
 \text{SRCO} = \frac{1}{\sqrt{n}} \Bigg|\Bigg|\text{ColSum} (\sqrt{\mathcal{C}})\Bigg{|}\Bigg |_2 -1 .
\end{align}
\normalsize

To discourage the model from collapsing onto homogeneous assignments, we take the element-wise square root of the community assignment matrix before applying the column sum. A simple demonstration of the effects of our procedure is shown in App. \ref{reg}. Given this, our full model can be fit end-to-end with the following cost function:
\small
\begin{align}
    \mathcal{L}_{\text{DMoN}} =-\underbrace{\frac{1}{2w} \text{Tr}(\mathcal{C}^\top B \mathcal{C})}_{\text{Modularity Loss}} + \underbrace{\frac{1}{\sqrt{n}} \Bigg|\Bigg| \text{ColSum}(\sqrt{\mathcal{C}})\Bigg|\Bigg|_2 -1}_{\text{Regularizer}}.
\end{align}
\normalsize
We use statistical process monitoring (SPM), a form of anomaly detection to monitor change in modularity. In general, SPM involves modeling a system's usual behavior and identifying abnormalities by prospectively tracking a key statistic, \(S_t\), in our case $Q_t$, the estimated modularity at a snapshot $t$. Monitoring occurs in two phases: Phase I sets control limits based on baseline behavior from sampled statistics, while Phase II monitors new values, flagging anomalies if they fall outside these limits.  Control charts plot these limits together with a time series of observations. Different control charts detect various changes, including Shewhart, cumulative sum (CUSUM), and exponentially weighted moving average (EWMA) charts. EWMA charts are favored for detecting gradual and sustained changes by modeling the exponentially weighted moving average, \(Z_t\), of \(S_t\):
\small
\begin{alignat}{5}
Z_t = \alpha S_t + (1-\alpha)Z_{t-1}
\end{alignat}
\normalsize
where \(0<\alpha<1\) is a smoothing constant, with \(Z_0\) often set as the estimated mean \(\hat{\mu} = \bar{S}\). Phase I data determines control limits:
\small
\begin{align}
\hat{\mu} \pm 3 \hat{\sigma} \sqrt{\frac{\alpha}{2- \alpha} [1- (1-\alpha )^{2t}]}
\end{align}
\normalsize
with \(\hat{\sigma}\) being the estimated standard deviation of \(S_t\).
The smoothing parameter \(\alpha\) is crucial for sensitivity; values closer to 1 emphasize recent observations, affecting detection rates. An \(\alpha\) of 0.2 is commonly used to balance gradual change detection and limit false alarms \cite{lucas_exponentially_1990}.  We note that we directly train our DMoN model on Phase I snapshots and reserve Phase II purely for evaluation. Our full procedure is visualized and further described in App. \ref{diag}



\section{Experiments}
\subsection{Data Generation}
We evaluate the performance of the proposed methodology with simulated experiments and synthetic data. Using the data generation technique from \citet{Synthetic}, we create data resembling real-world attributed graphs. This involves two steps: graph construction and attribute generation. First, we construct an initial graph using a degree-corrected stochastic block model (DCSBM) \cite{dcsbm}. The DCSBM models each edge weight via a Poisson distribution, with a mean depending on the nodes' community memberships and their propensity to connect. The probability mass function for the model is:
\small
\begin{equation}
\begin{aligned}
P(A|\boldsymbol{\theta},\Lambda,\mathbf{c}) &=  \prod_{u<v} \frac{1}{A_{uv}!} e^{-\lambda_{c_u,c_v}\theta_u\theta_v} (\lambda_{c_u,c_v} \theta_u \theta_v)^{A_{uv}} \\ 
 \times& \prod_{u} \frac{1}{(A_{uv}/2)!} e^{-\lambda_{c_u,c_u}\theta_u^2/2} (\lambda_{c_u,c_u} \theta_u^2/2 )^{A_{uv}/2}
\end{aligned}
\end{equation}
\normalsize

Here, $\boldsymbol{\theta}$ is the degree propensity vector, with $\theta_u$ being the degree propensity for node $u$, proportional to its expected degree. To ensure identifiability, the added normalization constraint  
    $\sum_u \theta_u \delta_{c_u,r} = 1$ must be applied.
$\Lambda$ is the $k\times k$ community propensity matrix, where an entry $\lambda_{r,s}$ indicates the edge propensity between communities $r$ and $s$. Here $\mathbf{c}$ is the community assignment vector, with an entry $c_u \in \{1,2,...,n\}$ indicating the community assignment of node $u$.  Node attribute information is generated from cluster-specific multivariate normal distributions, with cluster centers generated using a Gaussian mixture model. For each of the $s$ node attributes, we sample their attribute cluster centers from a $k$-dimensional multivariate Gaussian distribution with mean $\boldsymbol{\mu} = 0$ and covariance matrix $\Sigma = 3\times \boldsymbol{I}_{k\times k}$. Using the sampled cluster centers $\boldsymbol{\mu_c}$, we then sample the node attributes from an $s$-dimensional multivariate Gaussian distribution with covariance equal to the identity matrix. The simulation design includes 50 "in control" graphs for Phase I. After injecting a change point, 50 "out of control" graphs are monitored in Phase II. Each graph consists of $n = 1,000$ nodes in one of $k = 4$ equally sized communities. Node propensities, $\theta_u$, are generated using a uniform power law between $2^2$ and $2^6$. The intra-community propensity to connect is $\lambda_{s,s} = 18$, and the inter-community propensity is $\lambda_{r,s} = 2$ for $r,s \in \{1,2,...,k\}$.  Each node's attributes ($s = 64$) are sampled from their respective multivariate Gaussian distribution with mean $\boldsymbol{\mu}_c$ and covariance matrix $ \mathbf{I}_{s,s}$. This framework describes a single simulation run, we reapeat $N = 100$ times. Performance is evaluated using three metrics: change detection percentage, conditional expected delay, and average percentage of observations exceeding control limits. Change detection percentage is the proportion of runs in which a change was detected in at least one of the 50 out of control networks. Conditional expected delay is the average steps to detect the change, given detection occurs. The average percentage of the 50 out-of-control observations exceeding control limits, averaged over 100 runs, is also calculated. These metrics collectively provide insights into the method's change detection effectiveness and efficiency. 

\subsection{Large Changes}
We evaluate four changes to asses the efficacy of our method: Community splitting, community merging, new community, and attribute change.  \textit{Community Splitting}: Given the baseline parameters, we choose one of the four existing communities and randomly split its nodes into two new equally sized communities as well as sample new cluster centers, $\boldsymbol{\mu}_c$, for each of the two partitioned communities. We note this change does not affect the propensities to connect $\lambda_{r,r}$ and $\lambda_{r,s}$ remain unchanged. \textit{Community Merging}: to replicate real-life scenarios, our experimental set-up consists of merging all nodes of two communities into one. Additionally, we average the two community cluster centers to create the merged communities attribute centers. For example, the attribute center for the new community formed by merging communities $r$ and $s$ is $\boldsymbol{\mu}_{c,\text{merged}} = \frac{\boldsymbol{\mu}_{c,r} +\boldsymbol{\mu}_{c,s}}{2}$. \textit{New Community}: we model a sudden increase in nodes with unseen attributes by first increasing the number of nodes in the graph by $25\%$. Since our base scenario has $n = 1,000$ nodes, after the new community is inserted, the graph will have 1,250 nodes. Thereafter, we repeat the same procedure to generate the attribute information as with the first communities. \textit{Attribute change}: In contrast to sudden and abrupt changes, networks may also undergo gradual shifts in their underlying structure. We model such changes by periodically adding noise to the prior cluster centers $\boldsymbol{\mu}_c$. The process can be described by the following cluster center parameterization $
\boldsymbol{\mu_c}^{i+1} = \boldsymbol{\mu_c}^{i} + \boldsymbol{\tau}$
where $\boldsymbol{\tau}$ is a $s$-dimensional uniformly distributed noise term. That is, $\tau_j \sim U(0,1)$ for each $\tau_j \in \boldsymbol{\tau}$, and $i$ indicates the current iteration of the noise-adding process. Additionally, we note noise is only added during Phase II. Table \ref{CommunitySplittingTable} shows the evaluation for the reported changes. We observe our method is very effective at detecting these changes, with a change detection percentage of 1 and a very low expected delay. Additionally, we see that when the attributes change slowly, we are still able to detect this change reasonably quickly with a conditionally expected delay of 23.38 and a change detection percentage of 1.

\begin{figure}[t]
    \centering
    \includegraphics[width = .5\textwidth]{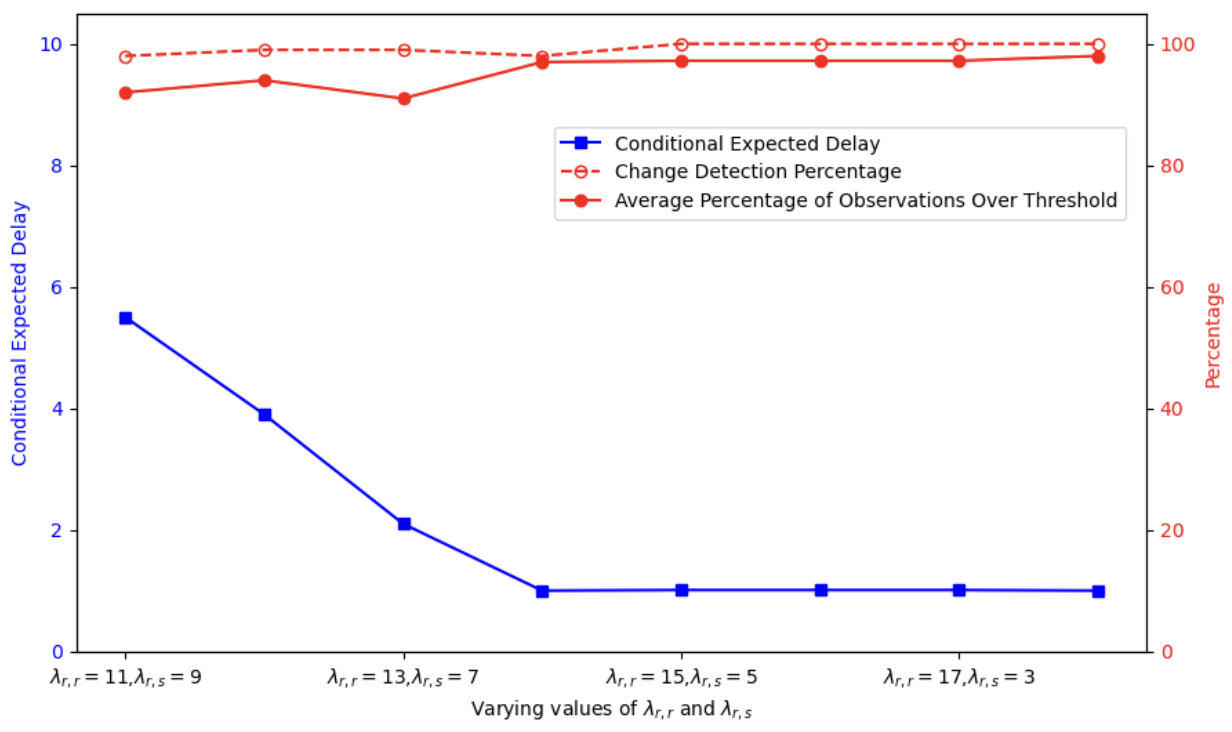}
    \vspace{-15pt}
    \caption{Results of synthetic experiments over varying community propensity, $\lambda_{r,r}$  and $\lambda_{r,s}$, values. Here each point represents $N=100$ simulations at their respective values of  $\lambda_{r,r}$  and $\lambda_{r,s}$}
    \label{fig:structure_control_final}
    \vspace{-15pt}
\end{figure}
\subsection{Structural Change}
Another important case is when node attributes are unchanged, but the graph generation process is altered. For this case, we consider imposing a change in the level of interaction between nodes by altering $\lambda_{r,r}$ and $\lambda_{r,s}$. We analyze a grid of changes in which we alter each pair of values sequentially as $\lambda_{s,r} = \lambda_{s,r} + 1$  and $\lambda_{s,s} = \lambda_{s,s} - 1$ and we repeat this process until $\lambda_{s,r} = \lambda_{s,s}$.  The resulting grid of $\lambda$ values are in the range of $\{10,...,18\}$ for $\lambda_{r,s}$ and $\{2,...,10\}$ for $\lambda_{s,s}$. We visualize how the inherent block structure changes in App. \ref{block}. Figure \ref{fig:structure_control_final} illustrates the performance of our method across these varying changes, each step illustrates $N=100$ simulations with those configurations.  We observe our method is quite accurate at detecting structural graph changes.  Although subtler structural changes, on average, require a few more timestamps, they are still detected fairly quickly. 

\begin{figure}[h]
\centering
    \includegraphics[width = .5\textwidth ]{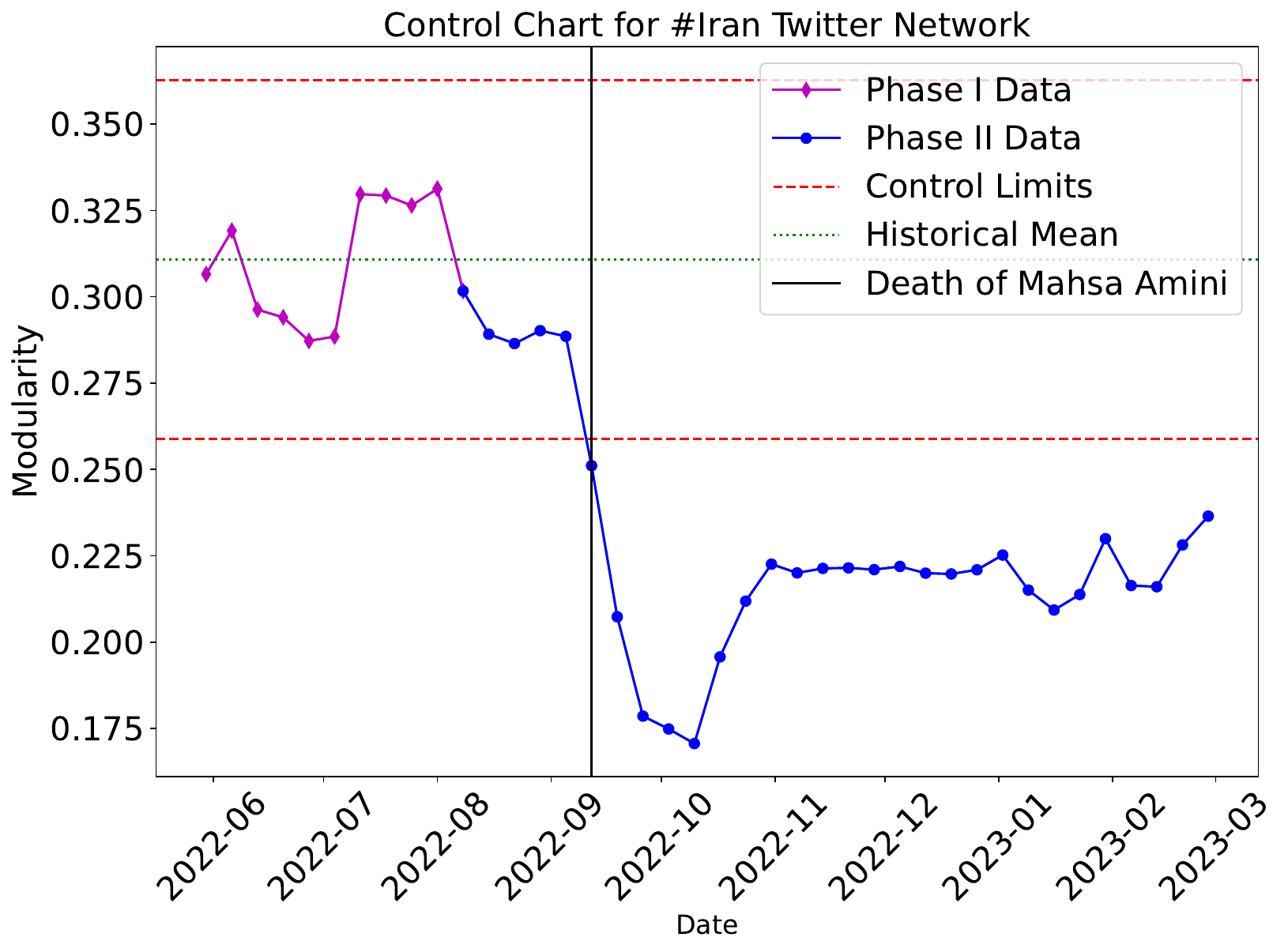}
    \vspace{-15pt}
    \caption{Control Chart for \# Iran Twitter Network when $\alpha =0.2$.}
    \label{fig:newControl}
    \vspace{-15pt}
    
\end{figure}
\subsection{Twitter Dataset}

In addition to simulation studies, we analyze a real-world Twitter dataset. On September 16, 2022, there was massive outrage within the Iranian community due to the death of Mahsa Amini, leading to increased activity in the \#Iran Twitter network \cite{ReliefWeb_2024_amani}. Here we explore whether our methodology can detect changes in networks during such real-world events. To determine this, we construct a network by using the Twitter API to extract all tweets containing the \#Iran hashtag. These tweets were collected over an eleven-month period and aggregated weekly; users are connected if they reply to each other. With these weekly snapshots, we build Phase I data from the first two months and Phase II with the rest. We use the text embeddings of user tweets obtained from a Sentence-Multilingual BERT model \cite{reimers2019sentencebert,conneau2018xnli} as node attributes. Figure \ref{fig:newControl} shows the control chart for this network. Clearly, the method detects a change point during the week of the incident. This demonstrates the practicality of our method when dealing with a real-world network with high-dimensional textual node attributes. We provide additional training and dataset details in App. \ref{Twitter}.

\section{Conclusion}
In this work, we propose a methodology for detecting community changes in highly attributed networks. Additionally, we derive the square root collapse regularizer, which penalizes suboptimal solutions previous regularizers failed to. Through extensive simulation studies, we showcase our method's ability to detect changes common in real-life networks. Additionally, we showcase the practicality of our method by analyzing a real-world network with textual node attributes. We demonstrate that our method is capable of detecting change associated with real-world events.

\nocite{langley00}

\bibliography{example_paper}
\bibliographystyle{icml2024}

\newpage

\appendix

\onecolumn
\section{Acknowledgements }
We note the first author of this work is a member of the Latinx community. They conducted all coding, ran all experiments and wrote the majority of this manuscript. 
\section{Related Work}

\label{related_work}
\subsection{Network Monitoring}

Network monitoring aims to identify both sporadic and ongoing changes in a dynamic network, and has become increasingly popular within the field of Network Science. One way to approach the problem is by extending process monitoring techniques commonly used in manufacturing and engineering  to the network setting (\cite{PastPresentNetwork}). \cite{FirstNetwork} were the first to coin the term network monitoring, and apply traditional statistical process monitoring techniques to networks. By surveying network summary statistics, such as betweenness, closeness, and density, they were able to identify changes in communication within the Al-Qaeda terrorist organization. There have been many further extensions of this work, where rather than simply tracking local graph statistics, the graph is assumed to follow a specified generative model whose parameter estimates can then be monitored. Some examples of such generative models are the degree corrected stochastic block model (DCSBM) (\cite{dcsbm}, \cite{NathanielDCSB}, \cite{newidk}), the hurdle block model (\cite{hurdleModel}) or the dynamic latent space model (\cite{dynamicLatent}) to name a few. Additionally, \cite{BayesianMethods} leveraged a Bayesian predictive distribution to define control limits to monitor levels of communication between individuals using either a Poisson or hurdle Poisson conditional distribution. 

While impactful, the mentioned methods lack the ability to account for covariate information, which is more often than not crucial to modeling the underlying system dynamics. Fortunately, there are works that take this into account, for example, \cite{LogisticMonitoring2} leverage a logistic regression model, which is able to account for node and/or edge covariates in order to determine the probability of an edge forming between two individuals. This methodology is convenient as they are able to directly monitor the model coefficients and attribute a change directly to a specific covariate. Likewise, \cite{hurdleMonitoring} leverage covariates through a hurdle model to explain the propensity of edge weights. Unfortunately, these methods are limited by the linearity of the underlying model, prioritizing interpretability over complexity. 

While using statistical process monitoring methods to survey changing networks is quite popular, many other methods exist to identify anomalies in such systems. \cite{AndyLaplacian} choose to monitor dynamic graphs by observing the behavior of the singular values of the graph Laplacian matrix. Likewise, \cite{MCMCChange}  tracks a fixed number of dyads over the observed snapshots. They model the distribution of dyads through a conditionally independent two-state Markov chain over a window of snapshots,  detecting a change point if the Kullback–Leibler distance between current and historical snapshots exceeds a threshold.

\subsection{Temporal Community Detection}

Temporal community detection has become an increasingly important problem over the past decade due to the rise of social-media, online marketing, and the overall increase in people's online presence. There are many ways to frame the problem of temporal community detection. \cite{labelPropagation} view temporal communities as a static entity, that is, a node's community label is detected over the proposed series of network snapshots and considered fixed over time. Under this view, they build the LabelRankT algorithm, which is a heuristic algorithm that allows similarly linked nodes to be assigned the same label. In contrast, \cite{InfiniteCom} proposes  viewing the evolution of network communities in a continuous fashion.

Although useful, these methods are highly sensitive to user assumptions such as number of communities, sensitivity to time decay, and underlying hyperparameter choices (\cite{temporalSurvey}).  Many works follow and extend the main ideas of the prior two methods, and similarly experience the same limitations.  \cite{ComEvolution}, generalize non-negative matrix factorization to build a community matrix that is able to be tracked and modeled through time but is highly sensitive to initial assumptions of the underlying community structure. \cite{Zhao2019} build a heuristic framework for evolving community detection that does not require an assumed initial number of communities, but its heuristic nature does not allow for a quantifiable way to measure the significance of the change.

Other than the mentioned problems, another common limitation of the aforementioned methods is their lack of covariate modeling. \cite{dualAttributed} consider this by building a generative model that takes into account two types of covariates: \textit{assortative} attributes, those associated with the individual (i.e., age, gender, etc.) and \textit{generative} attributes, those that influence a link formation (i.e., shared hobbies, occupation, etc). While this work is able to model the evolution of communities over time, it incorporates attributes through a logistic regression model, limiting it by the required assumptions of linear models.

\subsection{Deep Learning for Community Based Network Monitoring}

Deep learning has become an increasingly popular method when learning high dimensional data and has achieved state-of-the-art performance on text (\cite{NlpReview}), image (\cite{VisionReview}), and graph related tasks (\cite{GraphSurvey}). While not in direct application to networks, deep learning has even been used to monitor distribution shifts in dynamic systems, but has mainly been applied to raw time series (\cite{tsChange1},\cite{tsChange2}). Additionally, \cite{csTGN} developed a DL learning architecture that is able to detect changes to network communities over time and with the ability to adapt to changes in distribution through a meta-learning framework. Unfortunately, their work is rooted in a supervised learning framework that requires ground truth community labels, something that is very rare in real-world datasets. 

Unlike the aforementioned methods, there exists DL-based community detection methods that are not restricted by the need for ground truth labels.  For example, \cite{vgraph} assume edges are generated using a hierarchical procedure with a hidden variable representing node community membership. They then use a variational approach to reconstruct the adjacency matrix, concurrently finding the most likely community memberships. Similarly, \cite{dmon} propose deep modularity networks (DMoN) which find local graph partitions, synonymous with communities, by directly optimizing the spectral modularity statistic. While these models are able to find high-quality clusters, they operate on a single network and do not generalize to a set of time-varying networks.

\section{Regularizers}

\begin{figure}[h]
    \centering
    \includegraphics[width=150mm,scale=0.5]{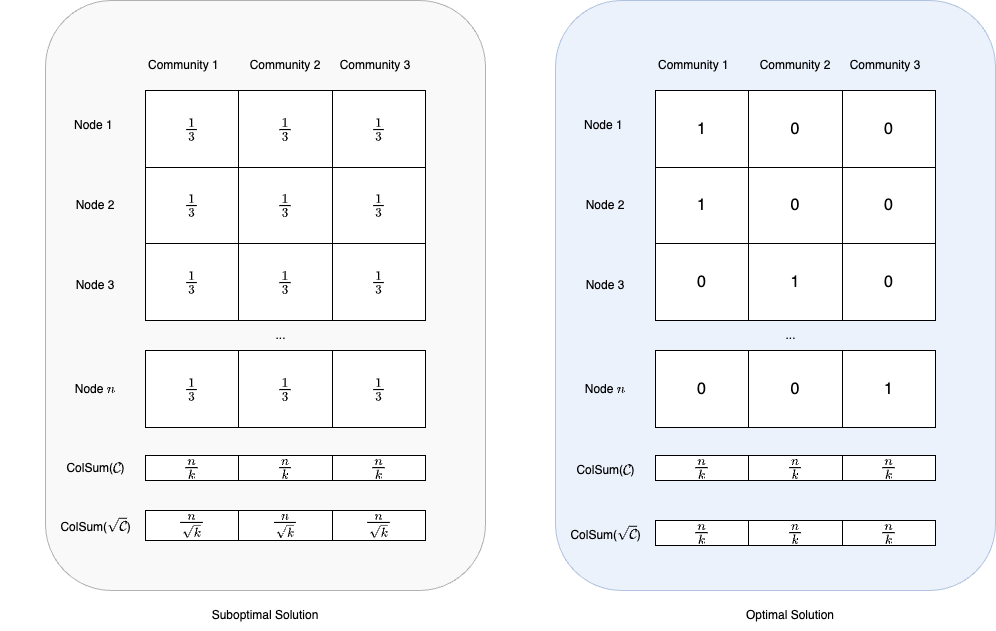}
    \caption{Comparison of square root collapse regularizer and collapse regularizer}
    \label{fig:Colapse-Regulizer}
\end{figure}

A simple demonstration of the effects of our procedure are shown in Figure \ref{fig:Colapse-Regulizer} which  compares our square root collapse regularizer and the collapse regularizer. For this illustration, we use a simple example with 3 evenly sized communities, $k=3$, but it generalizes to any $k \ge 2$. When employing the collapse regularizer, assigning a uniform distribution over the community assignments to all the nodes is equivalent to the optimal solution of assigning communities with certainty. On the other hand, the square root collapse regularizer addresses this issue by providing a larger value for the suboptimal solution, which can then be penalized.
\label{reg}

\section{Training Setup}

\begin{figure}[H]
    \centering
    \makebox[\textwidth]{  \includegraphics[width=110mm,scale=0.5]{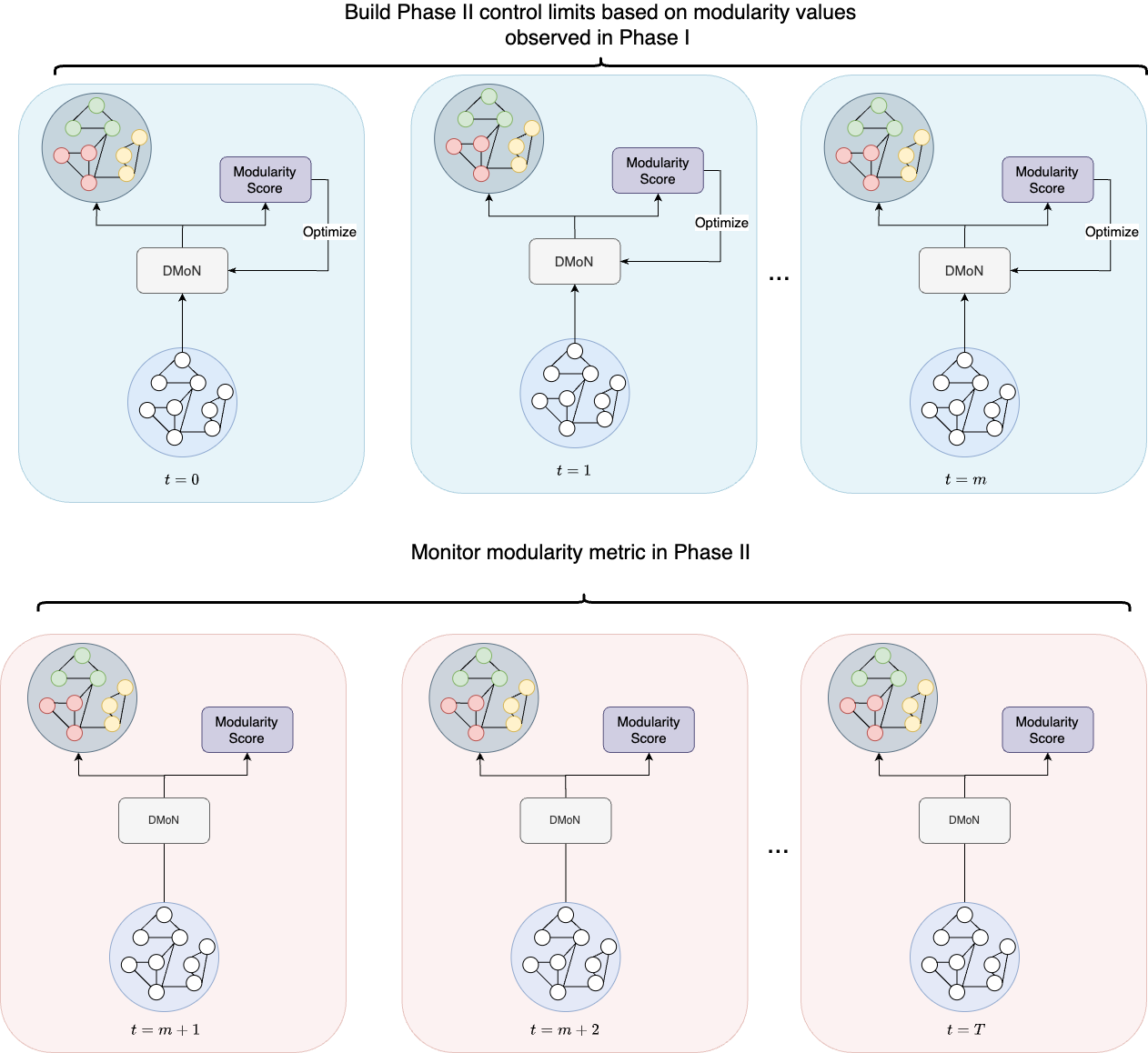}}%
    \caption{Illustration of model learning during Phase I $\&$ monitoring during Phase II}
    \label{fig:training}
\end{figure}

To be able to utilize a control chart with our method, we define Phase I by first training our model over a dynamic graph with snapshots, $G_1,...,G_m$. This allows our model to learn the stationary behavior of the network. Once converged, we calculate the modularity of the remaining unobserved snapshots, $G_{m+1},...,G_T$. In Phase II, we track the behavior of the graph using an EWMA control chart, where we choose $S_t$ to be the modularity scores of each successive graph. Through extensive synthetic experimentation, we demonstrate that this methodology is able to robustly detect changes in both graph attributes and graph structural changes.  Our methodology is illustrated in Figure \ref{fig:training}.

\label{diag}
\section{Block Visualization for Structural Changes}
\begin{figure}[H]
    \centering
    \makebox[\textwidth]{\includegraphics[width = 500pt]{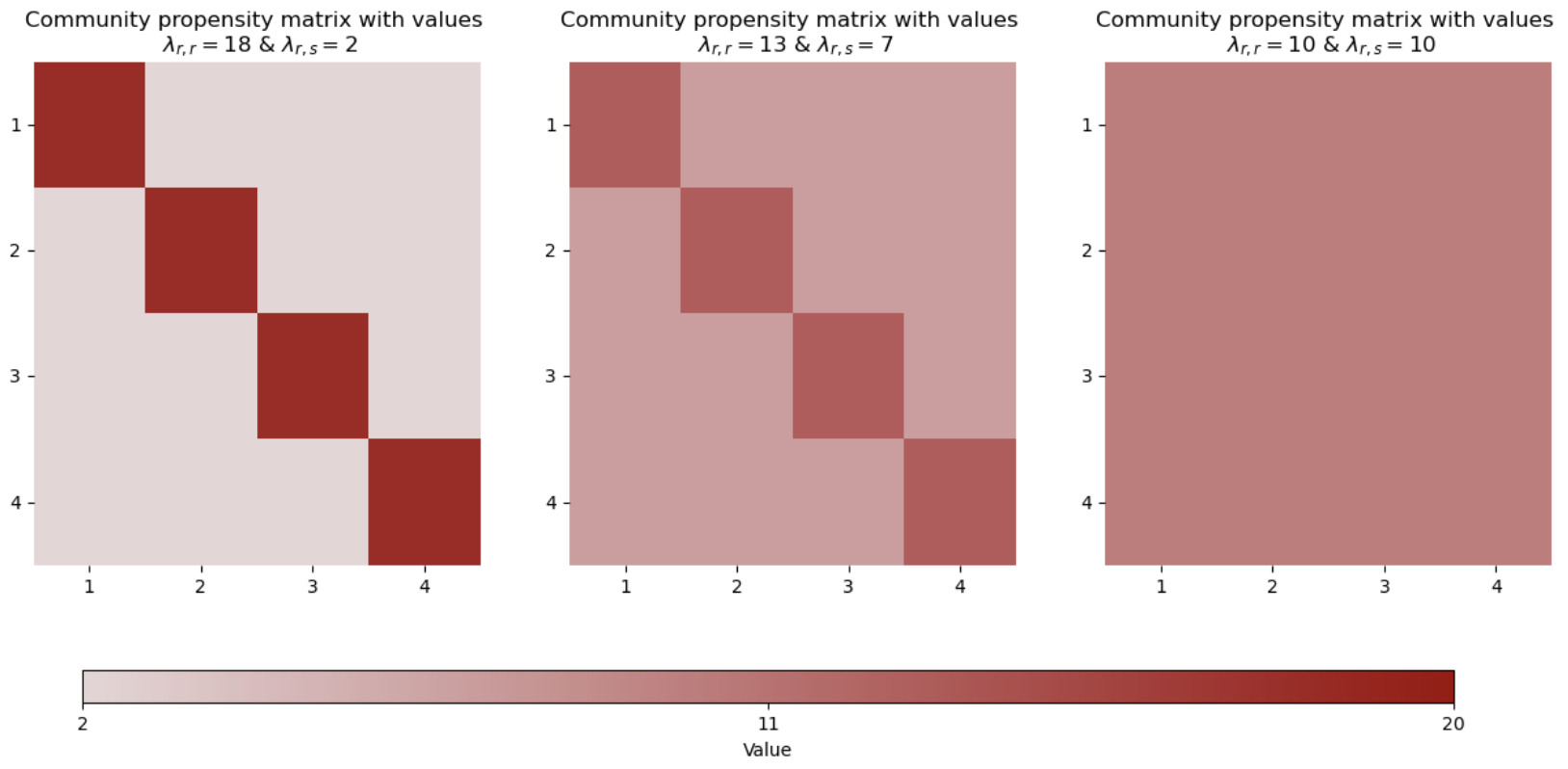}}%
    \caption{Heatmap representation of varying $\lambda_{r,r}$ and $\lambda_{r,s}$ values}
    \label{fig:blockHeatmap}
\end{figure}

Figure \ref{fig:blockHeatmap} illustrates how the $\Lambda$ matrix changes as we alter varying values of $\lambda_{r,r}$ and $\lambda_{r,s}$.
\label{block}

\section{No Change}
\label{no_chan}
\begin{figure}[h]
    \centering
    \includegraphics[width = 300pt]{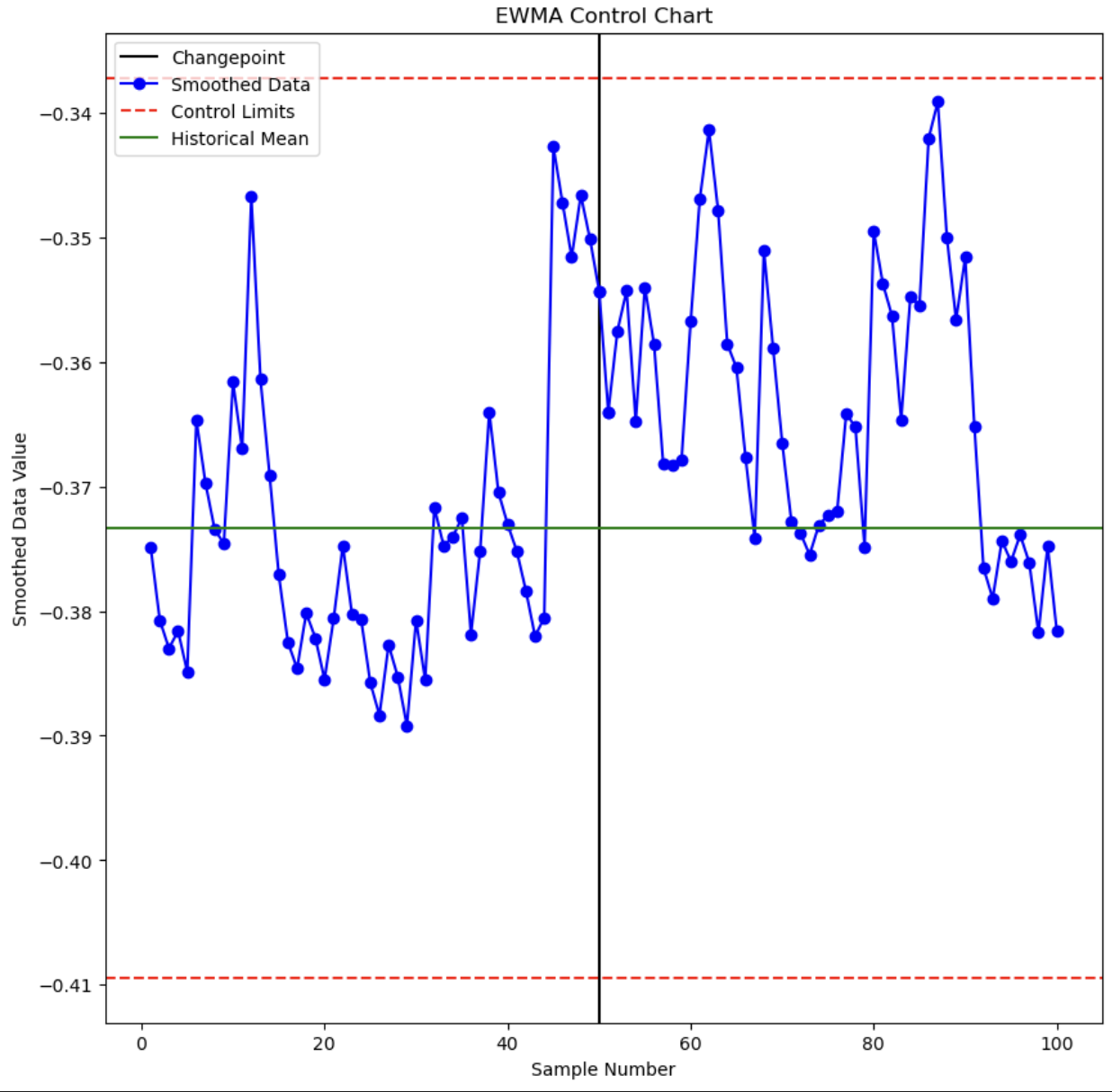}
    \caption{EWMA control chart when there is no change injected $\alpha =0.2$ }
    \label{fig:noChangeControlChart}
\end{figure}
In addition to the explored changes, our method must also be evaluated in scenarios where there is no change to the graph structure or attributes. Unlike the other sections focused on change detection, our objective in this case is different – we do not want to detect changes and hence having a low average percentage over the threshold becomes ideal as it suggests the robustness of our method to type I errors. The performance of our method under these conditions is illustrated in Table \ref{nochangeTable}, where we observe a consistently low average percentage over the threshold, indicating a low false alarm rate.  This is slightly larger than the nominal $0.05$ value; this small difference is likely due to the monitored modularity scores not precisely following the distributional assumptions made by the EWMA control chart. Figure \ref{fig:noChangeControlChart}, showcases an example control chart for this scenario. It can be seen there are no anomalous points in this case and the modularity remains stationary throughout both Phase I and II.

\begin{table}[h]
\centering
\begin{tabular}{|c|c|}
\hline
$\alpha$ & Average Percentage Over Threshold \\ \hline
0.2   & 0.07                                                           \\ \hline
\end{tabular}
\caption{Performance summaries across N = 100 replications when no change is injected }
\label{nochangeTable}
\end{table}

\section{\#Iran Network}
\label{Twitter}
\subsection{Data}
We scrape the \#Iran network for a seven-month period, an undirected edge is created when two users reply to each other. We build the text embedding by pooling the tweet embeddings of user tweets and their replies. If a user has multiple tweets in a single snapshot, we create a node embedding by averaging all tweets within that snapshot. We note the number of users may vary between each snapshot with an average of 5254.74 over all the snapshots with a standard deviation of 2,771.93. We collect the data over an eleven-month period from June 3rd 2022, to March 31st 2023. Unfortunately, due to Twitter policies, we cannot make this data public. However, we will open-source all code and simulation data upon acceptance.

\subsection{Training}
We train our DMoN model sequentially on Phase I graphs. We tune the number of communities ($k$), the learning rate, and dropout by reserving one Phase I snapshot for evaluation. Using the best performing model on the evaluation snapshot, we build the Phase I limits. We use this model throughout Phase II to calculate the modularity of each snapshot.  
\end{document}